\documentclass[12pt]{article}
\usepackage{cite,graphicx}

\def\beq{\begin{equation}}
\def\eeq{\end{equation}}
\def\bea{\begin{eqnarray}}
\def\eea{\end{eqnarray}}
\newcommand{\gsim}{\lower.7ex\hbox{$\;\stackrel{\textstyle>}{\sim}\;$}}
\newcommand{\lsim}{\lower.7ex\hbox{$\;\stackrel{\textstyle<}{\sim}\;$}}

\begin{document}
\thispagestyle{empty}
\noindent
DESY 03-158\hspace*{\fill} September 27, 2003\\
\vspace*{2.6cm}

\begin{center}
{\Large\bf Grand Unification in the Projective Plane}\\[2cm]
{\large A. Hebecker}\\[.5cm]
{\it Deutsches Elektronen-Synchrotron, Notkestrasse 85, D-22603 Hamburg,
Germany}
\\[1.5cm]

{\bf Abstract}\end{center}
\noindent
A 6-dimensional grand unified theory with the compact space having the 
topology of a real projective plane, i.e., a 2-sphere with opposite points 
identified, is considered. The space is locally flat except for two conical 
singularities where the curvature is concentrated. One supersymmetry is 
preserved in the effective 4d theory. The unified gauge symmetry, for 
example SU(5), is broken only by the non-trivial global topology. In 
contrast to the Hosotani mechanism, no adjoint Wilson-line modulus 
associated with this breaking appears. Since, locally, SU(5) remains a good 
symmetry everywhere, no UV-sensitive threshold corrections arise and 
SU(5)-violating local operators are forbidden. Doublet-triplet splitting can 
be addressed in the context of a 6d $N\!=\!2$ super Yang-Mills theory with 
gauge group SU(6). If this symmetry is first broken to SU(5) at a fixed 
point and then further reduced to the standard model group in the above 
non-local way, the two light Higgs doublets of the MSSM are predicted by 
the group-theoretical and geometrical structure of the model. 
\newpage

\section{Introduction}
The mechanism of gauge symmetry breaking is an important open issue in the 
context of grand unified theories (GUTs). In the conventional approach, 
where the symmetry is broken by the vacuum expectation values (VEVs) of GUT 
Higgs fields, large representations are usually required and a potential 
enforcing the desired VEV has to be specified. Furthermore, solving the 
doublet-triplet splitting problem without fine tuning adds extra complexity 
to the models. 

An interesting alternative is provided by the Hosotani mechanism~\cite{hos}, 
which can be implemented in higher-dimensional theories compactified on 
manifolds with non-trivial topology. In this case, the symmetry breaking can 
be ascribed to the VEV of a Wilson line wrapping a non-contractible loop in 
extra dimensions. Related geometrical mechanisms of gauge symmetry breaking 
are used in string-compactifications on Calabi-Yau manifolds~\cite{chsw} and 
can naturally lead to doublet-triplet splitting~\cite{wit}. Interesting 
relations between extra-dimensional topology and gauge symmetry breaking 
also exist in field-theoretic settings with non-zero field-strength and 
no supersymmetry (see, e.g.,~\cite{fm}).

A serious problem of field-theoretic GUT models with gauge symmetry breaking 
by the Hosotani mechanism is the flatness of the classical potential of the 
Wilson line, which is protected from loop corrections by supersymmetry 
(SUSY). Thus, one usually encounters light adjoint fields (with mass of the 
order of the SUSY breaking scale) ruining precision gauge coupling 
unification. 
The situation is improved in 5-dimensional field-theoretic orbifold GUT 
models~\cite{kaw} (see~\cite{dhvw} for the original stringy idea), where 
the gauge symmetry is broken by boundary conditions and no adjoint moduli
arise. Although related 6d GUT models~\cite{abc} again have Wilson line 
VEVs, these lines can be contracted to zero length at conical singularities 
of the compact space, so that the VEV is fixed by a symmetry-breaking 
boundary condition. The presence of such boundaries (locations with 
reduced gauge symmetry) restricts the predictivity of orbifold GUTs because 
fields and operators that do not respect the GUT symmetry can be added at 
these points. It also introduces UV-sensitive corrections to gauge coupling 
unification, the natural size of which corresponds roughly to the thresholds 
of conventional 4d GUTs.

In this letter, an alternative field-theoretic mechanism for gauge symmetry 
breaking is considered. To understand the basic idea, it is sufficient to 
consider a 6d field theory with gauge group $G$ compactified on a 2-sphere. 
Modding out by a $Z_2$ symmetry which acts on the sphere as a reflection 
with respect to the center and in gauge space by the inner automorphism 
$g\to PgP^{-1}$ (with $P\in G,\,\,P^2=1$), one obtains a gauge theory on the 
projective plane. The symmetry of the 4d effective field theory is reduced 
to the subgroup commuting with $P$. Alternatively, the model can be 
characterized as a sphere with the insertion of a crosscap (for the basic 
geometric concepts see, e.g.,~\cite{gsw}), where the identification of the 
opposite edges at the crosscap is associated with a gauge twist $P$. It is 
easy to observe (see also below) that the topology requires the gauge 
twist to obey $P^2=1$. Thus, although the breaking is entirely non-local,
no Wilson line modulus appears. More generally, this type of discretized 
topological breaking occurs in situations where the fundamental group of 
the compact space is non-trival but finite, such as in many Calabi-Yau 
models (see~\cite{chsw,wit} and, in particular,~\cite{ww}). In this context, 
the real projective plane has been mentioned in~\cite{hmn}. The present 
realization combines the features of an extremely simple compact space with 
unbroken $N\!=\!1$ SUSY and thus has all the ingredients necessary for 
realistic model building. 

The paper is organized as follows: In Sect.~\ref{bc}, it is shown how the 
above illustrative example can be promoted to a more interesting GUT-like 
model. In particular, starting from a 6d $N\!=\!2$ super Yang-Mills (SYM) 
theory, a model with unbroken 4d $N\!=\!1$ SUSY, broken gauge symmetry and 
no moduli is constructed by orbifolding. The geometry is such that the 
curvature of the topological 2-sphere discussed above is concentrated at 
four conical singularities. The construction involves modding out a freely 
acting discrete symmetry (cf. the freely acting orbifold models familiar 
in string theory, especially in the context of SUSY breaking~\cite{fao}). 

In Sect.~\ref{su6}, an SU(6) model with doublet-triplet splitting is 
discussed. Given that the smallest truly unified group is SU(5), it is 
desirable that no SU(5) breaking fixed points exist. 
Doublet-triplet splitting is then most naturally realized if the Higgs is a 
bulk field. Since no gauged bulk matter is allowed by 6d $N\!=\!2$ SUSY, 
the gauge group has to be extended to allow Higgs doublets to emerge from 
the adjoint representation, the minimal choice being SU(6). Indeed, it is 
possible to construct a model where SU(6) is broken to SU(5) at one of the 
fixed points (the further breaking being topological) and two naturally light 
Higgs doublets appear in a way closely related to the models 
of~\cite{hns,pst,bn}. 

Sect.~\ref{sing} is devoted to the fixed-point breaking of SU(6) to SU(5).
For this several possibilities exist, the simplest one being to declare 
the gauge symmetry at the conical singularity to be reduced and to supply 
appropriate boundary conditions for the bulk fields. A more interesting 
possibility involves cutting off the tip of the cone by a 
gauge-symmetry-breaking 5d boundary and letting the length of the 
boundary tend to zero.

Conclusions and open questions are discussed in Sect.~\ref{con}.

\section{Non-locally broken SYM theory without moduli}\label{bc}
Consider a 6d SYM theory with (1,1) SUSY and gauge group $G$, which can 
be thought of as deriving from a 10d SYM theory by torus compactification. 
(The large amount of supersymmetry is required to ensure the 
phenomenologically desirable $N=1$ SUSY of the 4d theory obtained 
after orbifolding.) It will prove convenient to describe this theory in 
terms of a 4d vector superfield $V$ and 3 chiral superfields with scalar 
components $\Phi_5=A_5+iA_8$, $\Phi_6=A_6+iA_9$ and $\Phi_7=A_7+i 
A_{10}$~\cite{mss}. Here $A_5$ and $A_6$ are the extra-dimensional (with 
respect to 4d) gauge field components of the 6d theory and $A_7\dots 
A_{10}$ are 6d adjoint scalars deriving in an obvious way from the gauge 
field of the associated 10d SYM theory.

The 6d theory is compactified on a torus $T^2$ parameterized by 
$(x_5,x_6)\in I\!\!R^2$ with the identifications $x_5\sim x_5+2\pi R_5$ and 
$x_6\sim x_6+2\pi R_6$. In the first step of orbifolding, the rotation 
symmetry $(x_5,x_6) \to -(x_5,x_6)$ is `modded out'. The action of this 
$Z_2$ symmetry in field space is chosen (using the symbol $\Phi_i$ for both 
the superfield and its scalar component) as $(\Phi_5,\Phi_6)\to -(\Phi_5, 
\Phi_6)$, with $V$ and $\Phi_7$ being inert. The fundamental compact space 
now has the topology of a 2-sphere with the curvature being concentrated 
at 4 conical singularities with deficit angle $\pi$ and can be visualized 
as the surface of a `pillow'~\cite{abc1}. The gauge symmetry is unrestricted. 

\begin{figure}[ht]
\begin{center}
\includegraphics[width=6cm]{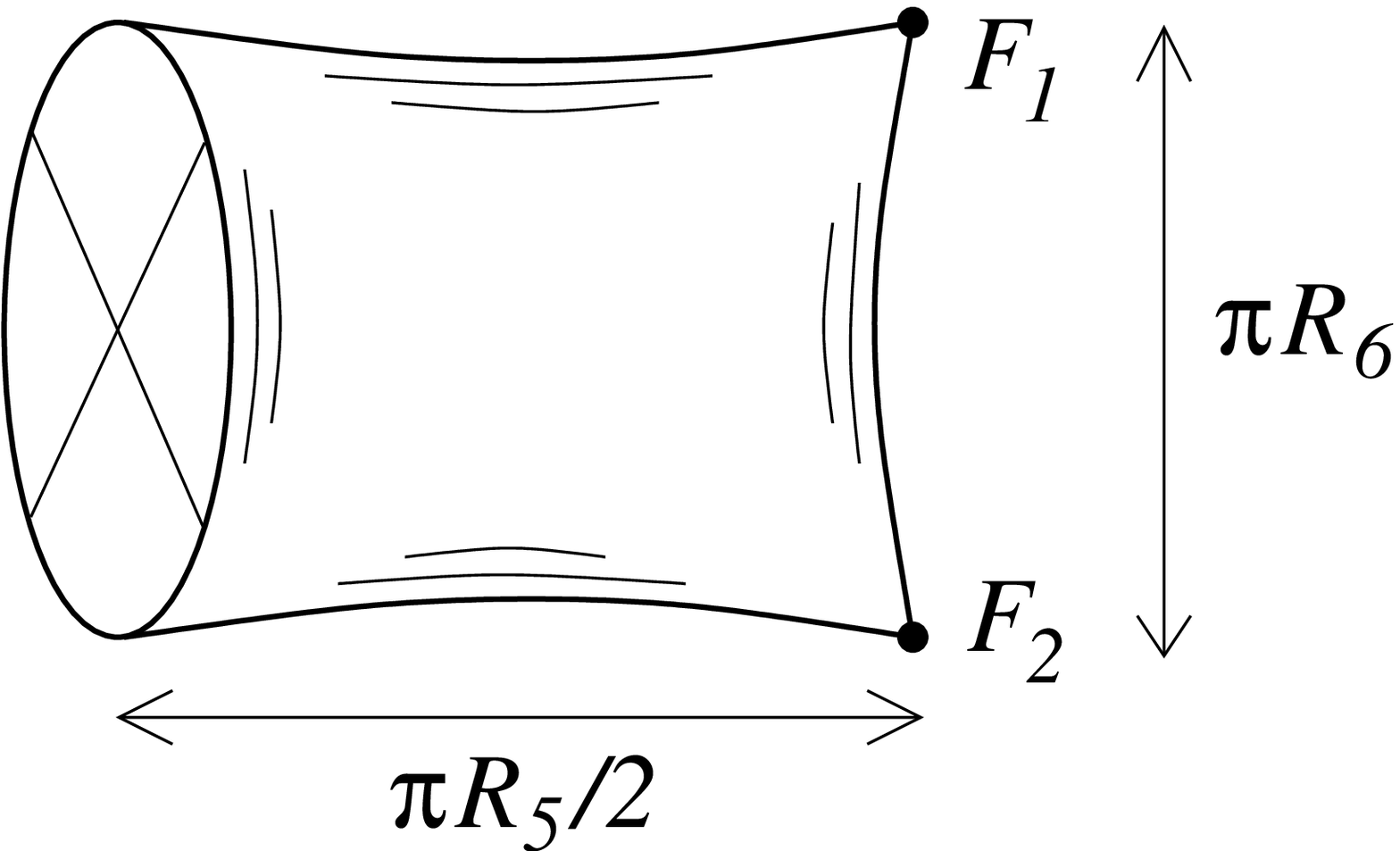}
\end{center}
\refstepcounter{figure}\label{cc}

\vspace*{-.2cm}
{\bf Figure~\ref{cc}:} Illustration of the topology of the $T^2/(Z_2\times 
Z_2')$ model discussed in the text. The manifold is flat everywhere except 
for the two conical singularities at $F_1$ and $F_2$. At the crosscap 
(symbolized by a circle with a cross) opposite points are identified, making 
the surface non-orientable.
\end{figure}

In the second, crucial step, a freely-acting $Z_2'$ symmetry is modded out,
at which point the gauge symmetry breaking is introduced.
The geometric $Z_2'$ action is defined by $x_5-\pi R_5/2\to 
-(x_5-\pi R_5/2)$ (reflection with respect to the line $x_5=\pi R_5/2$) and 
$x_6\to x_6+\pi R_6$ (translation along that line). The action in field 
space makes use of the gauge twist $P\in G$ ($P^2=1$) and is given by 
$V\to PVP^{-1}$, $\Phi_5\to -P\Phi_5P^{-1}$, $\Phi_6\to P\Phi_6P^{-1}$ and 
$\Phi_7\to -P\Phi_7P^{-1}$ (cf.~the rectangular models of~\cite{hns}). The 
topology of the resulting fundamental compact space (which is equivalent to
that of the projective plane) is illustrated in Fig.~\ref{cc}. The space 
is non-orientable, has no boundaries and the curvature is concentrated at the 
two fixed points $F_1$ and $F_2$, where conical singularities with deficit 
angle $\pi$ reside. 

Given the above $Z_2$ and $Z_2'$ action, it is clear that the massless modes 
are the fields from $V$ commuting with $P$ (corresponding to the unbroken 
subgroup $H\subset G$) and the fields from $\Phi_7$ anticommuting with $P$
(corresponding to the complement of $H$ in $G$). The latter ones can become 
heavy due to a mass term for $\Phi_7$ localized at one of the fixed points. 
One 4d SUSY survives since the non-gauge part of both the $Z_2$ and $Z_2'$ 
action on the $\Phi_i$ fits into SU(3)~\cite{chsw}. 

Thus, in the specific case $G=\,\,$SU(5) and $P=\,\,$diag$(1,1,1,-1,-1)$ 
(cf.~\cite{kaw}), the gauge symmetry is broken to the standard model (SM) 
group in an entirely non-local fashion (the fixed points are SU(5) 
symmetric) and only the SM gauge multiplet is naturally light. Although 
SM matter can be added at the two fixed points, it appears difficult to 
realize doublet-triplet splitting given the extremely soft nature of the 
breaking. By construction, the fixed points have 4d $N\!=\!2$ SUSY which 
can, however, be broken simply by allowing for operators or a field content 
that are consistent only with $N\!=\!1$. 

The most interesting feature of the above toy model GUT is the 
absence of the Wilson line moduli that typically accompany a purely 
topological gauge symmetry breaking. This effect is linked to the small 
fundamental group of the projective plane ($Z_2$ rather than, say, $Z$ for 
an $S^1$) and can be understood as follows: By construction, a closed Wilson 
line loop passing once through the crosscap has the value $P$. A Wilson loop 
passing through the crosscap twice has the value $P^2$ and is, at the same 
time, contractible by the topology of the projective plane. Thus $P^2=1$ is 
required by consistency as long as the field strength vanishes everywhere
including the fixed points (i.e., an infinitesimal loop surrounding a fixed 
point has value 1). In field theory, this latter statement corresponds 
simply to a gauge-invariant boundary condition at the singularity. The fact 
that such a boundary condition can not be continuously deformed is natural 
from the point of view of string theory (cf. the discrete or quantized 
Wilson lines of~\cite{dhvw,inq}). Changing the value of the non-contractible 
Wilson loop while its square is fixed to equal unity corresponds to a
change by inner automorphism and thus to an unphysical, gauge degree of 
freedom. 

Note that it is possible to give the model an equivalent formulation with 
trivial (from the gauge theory perspective) transition function at the 
crosscap and non-vanishing, smooth
gauge connection throughout the compact space. To see this, consider the 
same geometry as above and introduce a gauge field which, near the 
crosscap boundary, is pointed parallel to that boundary and has the same
value everywhere along that boundary. This value can be chosen such that a 
Wilson line loop through the crosscap equals $P$. Furthermore, let the field
strength be zero everywhere in the bulk and let the infinitesimal Wilson 
loop surrounding $F_2$ vanish. Then, by global topology (cf. the discussion 
of `conifold GUTs' in~\cite{hr}) the infinitesimal Wilson loop surrounding 
$F_1$ equals $P^2=1$ and thus gauge symmetry is unbroken at that point.

\section{An SU(6) model with doublet-triplet splitting}\label{su6}
As far as realistic model building is concerned, the most unsatisfactory 
feature of the above construction is the absence of doublet-triplet 
splitting. This problem can be overcome by extending the bulk gauge 
symmetry to SU(6) (cf.~\cite{hns,pst,bn}) and using the gauge twist 
$P=$\,\,diag$(1,1,1,1,-1,-1)$ for
the breaking at the crosscap. In addition, SU(6) is broken to SU(5) (defined 
as the submatrix containing the last 5 rows and last 5 columns of the 
original $6\times 6$ matrix) at the fixed point $F_1$. For the moment, 
the precise mechanism of this breaking can be left unspecified. (One may 
imagine boundary localized Higgs fields or a hard, ad-hoc breaking 
by boundary conditions.) It is, however, important that no non-trivial 
Wilson line surrounding $F_1$ is introduced and that the gauge fields in 
$V$ outside the SU(5) subgroup acquire mass due to local physics at $F_1$. 

Now, the low-energy gauge symmetry is that of the SM and the 
only potentially light fields (in addition to the gauge fields) are 
part of $\Phi_7$ and transform as 
\beq
({\bf 3},{\bf 2})_{-5}+({\bf\bar{3}},{\bf 2})_5+({\bf 1},{\bf 2})_{-3}+
({\bf 1},{\bf 2})_3
\eeq
under the SU(3)$\times$SU(2)$\times$U(1) subgroup of SU(5). They are 
associated with SU(5) multiplets relevant for the local physics at $F_1$:
\beq
({\bf 3},{\bf 2})_5+({\bf\bar{3}},{\bf 2})_{-5}\,\subset\,{\bf 24}\quad,
\qquad({\bf 1},{\bf 2})_3\,\subset\,{\bf 5}\quad,\qquad({\bf 1},{\bf 2}
)_{-3}\,\subset\,{\bf\bar{5}}\,.
\eeq
Phenomenologically, the fields with the quantum numbers of $X,Y$ gauge bosons 
have to become massive while the two pure doublets may stay light and play 
the role of the two SM Higgs fields. It is clearly possible to introduce 
extra symmetries at $F_1$ allowing the first $({\bf 24\times 24})$ and 
forbidding the second $({\bf 5\times\bar{5}})$ SU(5) invariant mass term. 
SM matter could than be added in the form of brane-localized fields at $F_1$. 
The bulk should not be too large so that non-local effects can lead to a 
sufficient violation of the SU(5) symmetry of the $F_1$-based Yukawa 
couplings. Except for the somewhat ad hoc (though not fine-tuned) lightness
of the Higgs doublets, the model now appears to be satisfactory. 

However, it turns out that the above analysis is incomplete and that the 
local breaking of SU(6) to SU(5) at $F_1$ leads to the appearance of extra 
light states, not visible in the usual Kaluza-Klein mode expansion. The 
point is that some of the degrees of freedom associated with the Wilson
loop going through the crosscap cease to be gauge artifacts and become 
physical fields. To be more specific, let the value of the closed Wilson 
loop beginning at $F_1$ and going once through the crosscap be $W\!\in\,\,
$SU(6). So far, we have assumed $W=P$. It will now be shown that changing $W$ 
by inner automorphism (e.g., to $W'=UPU^{-1}$ with $U\!\in\,\,$SU(6)$\,$) 
corresponds to the excitation of a physical degree of freedom if the gauge 
symmetry at $F_1$ is reduced. Before doing so, let us argue that this change 
of $W$ is indeed a flat direction in fields space. For this, it is 
sufficient to consider a smooth scalar function $f$ on the fundamental 
space (cf.~Fig.~\ref{cc}) such that $f\equiv 0$ in a neighbourhood of the 
crosscap and $f\equiv 1$ in a neighbourhood of $F_1$. Writing $U=\exp(T)$, 
it becomes clear that the gauge fields that would be introduced by a gauge 
transformation $\exp(fT)$ lead to the desired inner-automorphism change of 
$W$ while corresponding, at the same time, to a flat direction in field 
space. 

Now, it remains to be shown that some of the freedom of rotating $W=P$ to 
$W'=UPU^{-1}$ corresponds indeed to physical fields. Group-theoretically, the 
SU(6) breaking at $F_1$ may be thought of as coming from a $Z_2$ twist $P'\!=
\,\,$diag$(-i,i,i,i,i,i)$. Given the restricted gauge symmetry at $F_1$, it 
is possible to write down the gauge invariant (and thus observable) 
operator tr$(W'P')=\,$tr$(UPU^{-1}P')$. Clearly, excitations generated by 
Lie algebra elements $T$ that 
anticommute with both $P$ and $P'$ lead to a change of the value of this 
operator. Their quantum numbers are the same as those of the two SM Higgs 
doublets obtained above from the chiral superfield $\Phi_7$. However, their 
physical interpretation is entirely different. While the former are 
conventional bulk zero modes with potential brane-localized mass terms, the 
latter are non-local degrees of freedom parameterizing the relative 
orientation of the two breaking patterns $P$ and $P'$. Thus, no local 
mass term or couplings to other fields can be written down. However, such 
couplings may be generated non-perturbatively or by integrating out 
appropriate gauged bulk fields~\cite{hmn,cgm}. 

There are now various options for solving the doublet-triplet splitting 
problem: One can use the Higgs doublets from $\Phi_7$ and argue that 
effective non-local operators will make the Wilson-line doublets heavy. 
This means that the problem of Wilson-line moduli (avoided by the crosscap 
breaking) is partially reintroduced and then solved in an ad-hoc way.
Thus, it appears more natural to give all $\Phi_7$ fields brane masses and 
to view the light Wilson line doublets, the existence of which is deeply 
rooted in the geometry of the model, as natural candidates for light Higgs 
fields. With matter localized at $F_1$, Yukawa couplings now arise from 
non-local effects. The main potential problem of this scenario is the 
difficulty of generating a sufficiently large top mass. Finally, the 
situation could be more involved and the required two light Higgs fields
might arise as a non-trivial linear superposition of the four potentially 
light doublets. For the purpose of this paper, it is sufficient that the 
existence of natural solutions to the doublet-triplet splitting problem 
has been demonstrated. The analysis of possibilities for generating 
realistic Yukawa couplings is postponed to a further investigation.

\section{Breaking SU(6) to SU(5) at a singularity}\label{sing}
In the previous section, it was simply assumed that the gauge symmetry 
at $F_1$ is reduced to SU(5). Here, the corresponding possibilities for 
symmetry breaking are analysed in more detail and, in particular, a 
realization within the framework of field-theoretic orbifolding is suggested. 

Clearly, one of the options is symmetry breaking by a brane localized 
Higgs~\cite{nsw}. In this case, it is important to use a set of VEVs and 
appropriate extra symmetries ensuring that the ${\bf 24}$ from $\Phi_7$ 
acquires a brane mass while the ${\bf 5}$ and ${\bf\bar{5}}$ from $\Phi_7$ 
remain light. 

Within the framework of field theoretic orbifolding, the most natural 
choice appears to be a Wilson line with value $P'$ wrapping $F_1$. 
However, introducing such a Wilson line gives mass both to the doublets from 
$\Phi_7$ as well as to the Wilson line doublets discussed in detail in the 
previous section. Thus, doublet-triplet splitting becomes a serious problem. 

\begin{figure}[ht]
\begin{center}
\includegraphics[width=3.5cm]{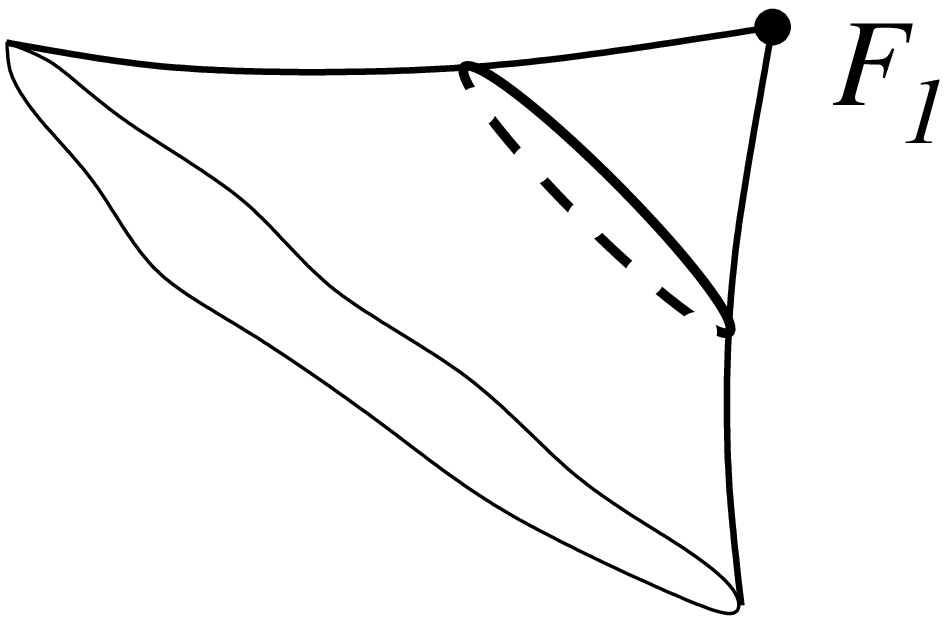}
\end{center}
\refstepcounter{figure}\label{cut}

\vspace*{-.2cm}
{\bf Figure~\ref{cut}:} Illustration of a possible resolution of the 
singularity at $F_1$. On taking the length of the cut to zero (i.e., 
cutting out a very small neighbourhood of $F_1$), the global structure of 
the original model of Fig.~\ref{cc} is recovered.
\end{figure}

Therefore, this section focusses on an alternative option where, as 
illustrated in Fig.~\ref{cut}, the singularity $F_1$ is cut out and removed 
from the fundamental space. The group SU(6) is broken by the $Z_2$ reflection 
used to define the boundary conditions at the 4+1 dimensional brane created 
by this cut (cf. the disc and annulus models of~\cite{li}). After cutting 
out $F_1$, the global topology is that of a M\"obius strip. One might be 
concerned that now moduli (corresponding to Wilson lines wrapping the cycle 
of the M\"obius strip) appear after all. However, they can be given a mass 
by appropriate non-local operators involving, e.g., the Wilson loop along 
the boundary. Taking the limit where the length of the boundary tends to 
zero, such operators become effectively local.

To be more specific about the boundary conditions, consider first a small 
interval on the boundary (implying that the curvature can be neglected) at a 
point $(x^5_0,x^6_0)$ where the boundary is parallel to the $x^6$ direction. 
One can then think of this boundary locally as being defined by the 
reflection $x^5-x^5_0\to -(x^5-x^5_0)$ together with an action in field 
space $V\to P'VP'^{-1}$, $\Phi_5\to -P'\Phi_5P'^{-1}$, $\Phi_6\to P'\Phi_6
P'^{-1}$ and $\Phi_7\to -P'\Phi_7P'^{-1}$. Choosing the boundary in the 
shape of a circle on the original torus, SO(2) rotation symmetry acting on 
the superfields $(\Phi_5,\Phi_6)$ as well as in coordinate space is now used 
to fix the boundary 
conditions locally at every point along the cut. In particular, although 
different linear combinations of $\Phi_5$ and $\Phi_6$ are forced to vanish
as one moves along the boundary, the fields from $\Phi_7$ anticommuting with 
$P'$ are always allowed to be non-zero. Thus, in the complete model, the 
$\Phi_7$ chiral superfield contributes precisely two potential Higgs 
doublets to the low-energy field content. This was realized at the 
beginning of the previous section by ad-hoc assumptions about local 
symmetry breaking and mass terms localized at $F_1$. Obviously, the breaking 
by $P'$ leads to a surviving U(1) in addition to SU(5). However, it does not 
affect the phenomenology at the present rough level of discussion and 
different options for the scale and mechanism of its breaking may be 
considered.

The above discussion of the $Z_2$ boundary condition ignored the 
curvature of the boundary in the flat 2d extra-dimensional space and 
treated it as a sum of straight elements. To describe the actual
curved boundary, one can simply go by diffeomorphism to a coordinate
system where the boundary is straight (and a corresponding non-zero Riemann
connection, which is considered non-dynamical, appears). Now a finite piece 
of the boundary can be defined by $Z_2$ reflection. In addition to the 
Riemann connection, an extra non-zero, non-dynamical R-symmetry connection 
has to be explicitly introduced in this picture. It accounts for the 
appropriate rotation of $A_8$, $A_9$ and the fermions in the $\Phi_5$ and 
$\Phi_6$ superfields as one moves along the boundary. 

\section{Conclusions}\label{con}
In this paper, gauge symmetry breaking by a specific topological feature, 
namely a crosscap, of the extra-dimensional manifold was considered. The 
main attributes of this simple mechanism are non-locality (implying the 
`softness' of the breaking), absence of flat directions (in particular 
Wilson line moduli), and the ease with which it can be implemented in a 
higher-dimensional SUSY GUT framework. 

If, in addition to the crosscap breaking, the gauge symmetry is broken 
by a different mechanism at some point in extra-dimensional space, 
Wilson-line moduli (characterizing the relative orientation of the two
breakings) reappear. For a breaking pattern leading from SU(6) to SU(5) 
and further to the standard model, this implies the existence of two 
light Higgs doublets.

In the context of the above SU(6) model, the main open questions concern 
the generation of Yukawa couplings and the possible role of two further 
potentially light Higgs doublets not associated with Wilson lines. In a
wider context, it would be interesting to consider larger groups, to
attempt to generate matter and Yukawa couplings from the pure SYM 
theory~\cite{bn,hs}, and to use the softness of the breaking to 
perform a calculation of GUT threshold corrections. Furthermore, a 
systematic understanding of other geometries (possibly in more than 2 
extra dimensions) allowing for this type of non-local but quantized gauge 
symmetry breaking would be desirable. 

\vspace*{-.3cm}
\section*{Acknowledgments}

\vspace*{-.3cm}
I would like to thank Wilfried Buchm\"{u}ller and Michael Ratz for useful 
discussions and comments.

\end{document}